\documentstyle[12pt]{article}
\begin{document}
\begin{titlepage}
\title{
{\bf Geometrical string and spin systems}
}
{\bf
\author{
 G.K. Savvidy\\
 Physics Department, University of Crete \\
 71409 Iraklion, Crete, Greece\\
 and\\
 Institut f\"ur Theoretische Physik \\
 D-6000 Frankfurt am Main 11,Germany\vspace{1cm}\\
 F.J.Wegner\\
 Institut f\"ur Theoretische Physik\\
 Ruprecht-Karls-Universit\"at Heidelberg\\
 Philosophenweg 19, D-69120 Heidelberg, Germany
}
}
\date{}
\maketitle
\begin{abstract}
\noindent

We formulate the geometrical string which has been proposed in
the articles \cite{savvidy1,savvidy2,savvidy3} on the euclidean
lattice. It is possible to find such spin systems with
local interaction which reproduce the same surface dynamics.
In the three-dimensional case this spin system is
a usual Ising ferromagnet with
additional diagonal antiferromagnetic interaction and with specially
adjusted coupling constants. In the four-dimensional case the spin
system
coincides with the gauge Ising system \cite{wegner} with an
additional
double-plaquette interaction and also with specially tuned
coupling constants. We extend this construction to random walks
and random hypersurfaces
(membrane and p-branes) of high dimensionality.
We compare these spin systems with the eight-vertex model and
BNNNI models.
\end{abstract}
\thispagestyle{empty}
\end{titlepage}
\pagestyle{empty}

\section{Introduction}
\vspace{.5cm}

 In the articles \cite{savvidy1,savvidy2,savvidy3}
the authors suggest a new
geometrical string, which can be considered as a natural extension
of the Feynman integral over paths to an integral over surfaces
in the sense, that
both amplitudes coincide in the cases, when the surface
degenerates
into a single particle world line. This geometrical string has
been formulated in continuum euclidean space $R^{d}$, where the
Feynman integral coincides with the partition function for the
randomly fluctuating surfaces.

In this article we formulate this surface model on a euclidean
lattice, where the surface is associated with the collection of
plaquettes, and we find statistical systems with local interaction,
which reproduce the same surface dynamics.

 In continuum euclidean space $R^{d}$
the random surface is associated with a connected polyhedral
surface $M$ with vertex coordinates
$X_{i}$, where $i = 1,..,\vert M \vert$ and $\vert M \vert$ is the
number
of the vertices. The energy functional for the polyhedral surface is
defined
as \cite{savvidy1}

$$A(M) = \sum_{<i,j>} \vert X_{i} - X_{j} \vert
\cdot \Theta(\alpha_{i,j}),\eqno(1)$$
where

$$\Theta(\pi) = 0 ,$$
$$ \Theta(2\pi -\alpha) = \Theta(\alpha),$$
$$\Theta(\alpha) \geq 0 ,\eqno(2)$$
the summation is over all edges $<i,j>$ of $M$ and $\alpha_{i,j}$
is the angle between the two neighbor faces
(flat polygons) of $M$ in $R^{d}$
having a common edge $<i,j>$.

 For triangulated surfaces one can define the
partition function

$$Z(\beta) = \sum_{M \in \{ M \} } \int e^{-\beta A(M)}
 \prod_{i \in M} dX_{i}, \eqno(3)$$
where $\{ M \}$ denotes a set of triangulations. The physical
properties of the model have been discussed in
\cite{savvidy1,savvidy2,savvidy3}.

 In the present article we formulate this string
on a euclidean lattice and find a corresponding lattice
field theory to which it is equivalent. A priory it is not evident
that an equivalent field theory exists. Particularly
we will show, that for euclidean
lattice surfaces, built from plaquettes, there exists a spin
system whose interface energy coincides with the expression (1).

 This will be done in two steps. In sect. 2 we will
formulate the lattice version of the string proposed in
\cite{savvidy1,savvidy2,savvidy3},
and in sect. 3 we will construct an appropriate spin system with
local interaction. The general formulation for $d-n$-dimensional
hypersurfaces in $d$-dimensional space is given in sect. 4
\vspace{1cm}

\section{Surfaces on a lattice}
\vspace{.5cm}

 On the lattice a closed surface can be considered as a
collection of plaquettes, whose edges are glued together
pairwise. There are two essentially distinct cases. In the
first case the surface is considered as a connected, orientable
surface with given topology and it is assumed, that
self-intersections of the surface do not
produce any additional energy. The surface should be allowed
to be freely intersected.
In this non-self-avoiding case the definition of the surface energy,
as it is defined by the expression (1),
is complete.

 In the second case it is assumed, that
self-intersections of the surface produce an additional energy and
one should define nontrivial weights associated with these
intersections \cite{savvidy3}.
This "fermionic" case corresponds to an
effective soft-self-avoidance
of the surface, and the corresponding weights should be defined as
\cite{savvidy3}

$$ \sum_{common~edges} \vert X_{i} - X_{j} \vert
\cdot (\Theta(\alpha^{(1)}_{i,j}) +...
+\Theta(\alpha^{(r(2r-1))}_{i,j}) ),\eqno(4a)$$
where $\alpha^{(1)}_{i,j},..,\alpha^{(r(2r-1))}_{i,j}$ are the
angles between the pairs of plaquettes of $M$
at the common edge $<i,j>$ in $R^{d}$ in which $2r$ plaquettes
intersect. The order of self-intersection is $r = 0,1,..,d-1$.

 On the lattice the lengths of the elementary edges
$ \vert X_{i} - X_{j} \vert $ are equal to the lattice constant $a$ and
the angles between plaquettes are either $0$,$ \pi /2$ or $\pi$.
Therefore we should only define quantities $\Theta(0)$,
$\Theta(\pi /2)$ and $\Theta(\pi) $ in (1), where the last one is
actually equal to zero (2).
In the case of infinitely
heavy quarks, that is $\Theta(0) = \infty$
\cite{savvidy2,savvidy3},
one can
neglect fluctuations of the surface with overhanging
plaquettes, that is with folds ( $\alpha_{i,j} = 0$).
Therefore only $\Theta(\pi /2)$
remains as a free parameter of the theory.

Thus we have a full
set of rules which allow to compute the energy functional
and to define the surface
dynamics on the lattice.
These rules are:
i) if two plaquettes of the closed surface
intersect under the right angle, the contribution
to the energy is equal to $a \cdot \Theta(\pi /2)$,
ii) if they are parallel, the contribution is equal to
zero $a\cdot \Theta(\pi) = 0$ and
iii) if four or more plaquettes intersect on a given edge,
then $a\cdot \Theta(\pi)$
has to be multiplied by the number of pairs of plaquettes
which meet under a right angle. Finally
the full energy associated with the lattice surface can
always be written as

$$H = \sum_{over~all~edges} H_{edge}, \eqno(5a)$$
where $H_{edge}$ is the energy associated with a given edge.
This completely defines the surface dynamics on the lattice.

In the following we will set $a=1$ and in many cases we will
suppress the coupling $\Theta(\pi /2)$.
\vspace{1cm}

\section{Equivalent spin model}
\vspace{.5cm}

Now let us attach variables $U_{P}$ to each
plaquette $P$ of the lattice \cite{wegner}.
By definition its value is $-1$, if a given plaquette belongs to the
surface $M$ and it is equal to $+1$ otherwise,

$$U_{P} = -1 \eqno(6a)$$
if $P \in M$ and
$$U_{P} = +1 . \eqno(6b)$$
if $P \not\in M$.
Since only an even number of plaquettes can meet
at a given lattice edge, the total
product of corresponding plaquette
variables $U_{P}$ surrounding an edge is always equal to one,

$$U_{1} U_{-1} U_{2} U_{-2} = 1 \eqno(7a)$$
if $d=3$ and
$$U_{1} U_{-1} U_{2} U_{-2} U_{3} U_{-3}=1 \eqno(7b)$$
if $d=4$, where $U_{P}$ denotes
corresponding plaquette variables and we choose
the given edge in the 3-rd direction when d=3 and in
the 4-th direction when d=4.

For our set of rules
the energy of the lattice surface
can be reexpressed through the plaquette variables $U_{P}$.
The energies of all possible configurations at a given lattice edge is
given by (we assume $a=1$)

$$H_{edge} = \frac{1}{4} (2-U_{1}-U_{-1})(2-U_{2}-U_{-2})
\Theta(\pi /2), \eqno(8a)$$
if $d=3$ and by
$$ = \frac{1}{4}(2-U_{1}-U_{-1})(2-U_{2}-U_{-2})\Theta(\pi /2)
\eqno(8b)$$
$$+ \frac{1}{4}(2-U_{2}-U_{-2})(2-U_{3}-U_{-3})\Theta(\pi /2)$$
$$+ \frac{1}{4}(2-U_{1}-U_{-1})(2-U_{3}-U_{-3})\Theta(\pi /2),$$
if $d=4$ .

 In three- and four-dimensions these formulae
together with (5), define the full Hamiltonian
of the desired system.
As it follows from constraints
(7), the local plaquette variables $U_{P}$ are not
completely independent. To
resolve constraints and to describe the system in terms of
independent local variables we should introduce Ising spins
on the dual lattice \cite{wegner}.

 In three dimensions, $R^{3}$, one should attach spin variables to
vertices of the dual lattice and represent plaquette variables
$U_{P}$ (7a) in the form
$$U_{1} = \sigma_{1}\sigma_{2}, U_{-1} = \sigma_{-1}\sigma_{-2},
U_{2} = \sigma_{1}\sigma_{-2}, U_{-2} =
\sigma_{-1}\sigma_{2}, \eqno(9)$$
where $\sigma_{r} = \pm 1$ denotes four Ising spin variables
surrounding
a given edge.
This representation resolves the constraints (7a) and yields

$$H_{edge} = 1-\frac{1}{2}(\sigma_{1}\sigma_{2} +
\sigma_{2}\sigma_{-1}
+\sigma_{-1}\sigma_{-2} + \sigma_{-2}\sigma_{1})
+ \frac{1}{2}(\sigma_{1}\sigma_{-1} + \sigma_{2}\sigma_{-
2}).\eqno(10)$$
The full Hamiltonian is the sum
of $H_{edge}$ over all elementary lattice edges (5)

$$H = -2J\sum_{\vec r,\vec \alpha}\sigma_{\vec r}
\sigma_{\vec r+\vec \alpha} +
\frac{1}{2}J \sum_{\vec r,\vec \alpha,\vec \beta}
(\sigma_{\vec r}\sigma_{\vec r+\vec \alpha+\vec \beta} +
\sigma_{\vec r+\vec \alpha}\sigma_{\vec r+\vec
\beta})\eqno(11)$$
The vector $\vec \alpha$ runs over the $d$ unit vectors parallel to
the $d$ axes. Similarly the sum over $\vec \alpha$ and $\vec
\beta$ runs over different pairs of such vectors.
The Hamiltonian represents a magnetic system with competing
interaction and specially adjusted coupling constants

$$ J_{ferromagnet} = 4J_{antiferromagnet} \equiv 2\Theta(\pi /2)
\eqno(12a)$$
Hamiltonian (11) coincides with usual Ising ferromagnet
with additional diagonal
antiferromagnetic interaction.
As it is easy to see from (11), here the ground state is more
degenerate than in the usual Ising case.

In order to resolve the constraints (7b) in four dimensions, one
should
attach independent spin variables $\sigma(r)$ to links of the
dual lattice (or what is equivalent at the center of plaquettes of
the initial lattice)\cite{wegner}. This allows to represent
plaquette variables (7b) in the form of product of spins
around the dual plaquette

$$U_{1}= \sigma_{2}(r)\sigma_{3}(r)
\sigma_{2}(r-e_{3})\sigma_{3}(r-e_{2}),~~~~~~~~$$
$$U_{-1}= \sigma_{2}(r-e_{1})\sigma_{3}(r-e_{1})
\sigma_{2}(r-e_{3}-e_{1})\sigma_{3}(r-e_{2}-e_{1}),$$
$$U_{2}= \sigma_{1}(r)\sigma_{3}(r)
\sigma_{1}(r-e_{3})\sigma_{3}(r-e_{1}),~~~~~~~~$$
$$U_{-2}= \sigma_{1}(r-e_{2})\sigma_{3}(r-e_{2})
\sigma_{1}(r-e_{3}-e_{2})\sigma_{3}(r-e_{1}-e_{2}),$$
$$U_{3}= \sigma_{1}(r)\sigma_{2}(r)
\sigma_{1}(r-e_{2})\sigma_{2}(r-e_{1}),~~~~~~~~$$
$$U_{-3}= \sigma_{1}(r-e_{3})\sigma_{2}(r-e_{3})
\sigma_{1}(r-e_{2}-e_{3})\sigma_{2}(r-e_{1}-e_{3}),\eqno(13)$$
Together with (8b) this yields

$$H_{edge}= 3 - \sum_{6~plaquettes}
(\sigma\sigma\sigma\sigma) + \frac{1}{4}
\sum_{12~right~angle~plaquettes}
(\sigma\sigma\sigma\sigma_{\alpha})^{rt}
(\sigma_{\alpha}\sigma\sigma\sigma).\eqno(14)$$
The full gauge invariant
Hamiltonian in four dimensions reads

$$H= -\frac{4}{g^{2}} \sum_{plaquettes}
(\sigma\sigma\sigma\sigma) + \frac{1}{4g^{2}} \cdot
\sum_{right~angle~plaquettes}
(\sigma\sigma\sigma\sigma_{\alpha})^{rt}
(\sigma_{\alpha}\sigma\sigma\sigma),\eqno(15)$$
where $g^{2}$ is the gauge coupling constant and

$$J_{plaquettes}= 16\cdot J_{rt~plaquettes} =
\frac{4}{g^{2}}.\eqno(12b)$$
Now one can define the partition function as

$$Z(\beta) = \sum_{\{\sigma_{\vec r}\}} e^{-\beta H}. \eqno(16)$$
Thus we have obtained the equivalent Ising models in three and
four dimensions.


\section{Model in higher dimensions}
\vspace{.5cm}

Equivalent spin systems can also be introduced in high
dimensions. For this purpose we introduce
an appropriate condense notation. It is also possible to
extend this construction to random walks and random
hypersurfaces of high dimensionality (membranes and p-branes),
when the energy functional has the same
nature (1),(4).

The points of the lattice are given by a $d$-
dimensional vector $\vec r$ whose components $r_{1},..,r_{d}$
are integer. In this lattice, $d-n$ dimensional elementary
hyperplaquettes $\Omega_{\alpha_{1}...\alpha_{n}}(\vec r)$, (all
$\alpha_{i}$
are different) are defined by $x_{\alpha_{i}} = r_{\alpha_{i}}$
and $r_{\alpha} \leq x_{\alpha} \leq r_{\alpha}+1$ for all
other $\alpha$. The $d-n-1$ dimensional
boundaries of the hyperplaquette
$\Omega_{\alpha_{1}...\alpha_{n}}(\vec r)$
are the $2(d-n)$ hyperedges
$\Omega_{\alpha_{1}...\alpha_{n}\beta
}(\vec r)$
and $\Omega_{\alpha_{1}...\alpha_{n} \beta}(\vec r+\vec
e_{\beta})$,
where $\beta$ differs from the $\alpha_{i}'s$ and $\vec e_{\beta}$
is the unit vector in $x_{\beta}$ direction.

A $d-n$ dimensional hypersurface $M_{d-n}$ in this lattice
consists of a collection of the
elementary hyperplaquettes
$\Omega_{\alpha_{1}...\alpha_{n}}(\vec r)$.
As before we introduce a variable
$U_{\alpha_{1}...\alpha_{n}}(\vec r)$, which assumes the value
$-1$ if the hyperplaquette belongs to the hypersurface $M_{d-n}$
and
$+1$ if it does not. The hypersurface $M_{d-n}$ is closed, if
each $d-n-1$ dimensional hyperedge
$\Omega_{\alpha_{1}...\alpha_{n+1}}(\vec r)$ belongs totally to an
even
number of $d-n$ dimensional hyperplaquettes of hypersurface
$M_{d-n}$.

Now we can introduce an energy of the closed $d-n$
dimensional hypersurface. This energy is a sum over all $d-n-1$
dimensional hyperedges

$$H = \sum H_{\alpha_{1}...\alpha_{n+1}}(\vec r). \eqno(5b)$$
The contribution of such a hyperedge depends on the $U$'s
of the surrounding $d-n$ dimensional hyperplaquettes. The energy
can always be expressed by a polynomial in $U$'s due to the
identity

$$H(\{U\}) = \sum_{U'}\prod_{i}\frac{1+U_{i}U'_{i}}{2}
H(\{U'\}).\eqno(17)$$

If the energy is the area of the hypersurface $M_{d-n}$, that is the
number of $d-n$ dimensional elementary hyperplaquetts, then we
can write it as \cite{wegner}

$$H^{area} = \sum\frac{1-U_{\alpha_{1}...\alpha_{n}}(\vec r)}{2}.
\eqno(18)$$
This is the type of Hamiltonians considered by Wegner
\cite{wegner}.
Since each $d-n$ dimensional hyperplaquette belongs to the $2(d-
n)$
hyperedges $\Omega_{\alpha_{1}...\alpha_{n}\beta }(\vec r)$ and
$\Omega_{\alpha_{1}...\alpha_{n} \beta}(\vec r+\vec e_{\beta})$,
we may
write the contribution of $\Omega_{\alpha_{1}...\alpha_{n+1}}(\vec
r)$
in the area case as
$$H^{area}_{\alpha_{1}...\alpha_{n+1}}(\vec r) = $$
$$\frac{1}{4(d-n)}
\sum_{k}(2- U_{\alpha_{1}...
\alpha_{k-1}\alpha_{k+1}...\alpha_{n+1}}(\vec r) -
U_{\alpha_{1}...\alpha_{k-1}\alpha_{k+1}...
\alpha_{n+1}}(\vec r-\vec e_{\alpha_{k}})).\eqno(19)$$

Instead of that we wish to introduce an energy which counts at
each
hyperedge $\Omega_{\alpha_{1}...\alpha_{n+1}}(\vec r)$
the number of pairs of
hyperplaquettes $\Omega_{\alpha_{1}...
\alpha_{k-1}\alpha_{k+1}...\alpha_{n+1}}$
that meet under a right angle. Apparently this is given by

$$H_{\alpha_{1}...\alpha_{n+1}}(\vec r) = $$
$$\frac{1}{4}
\sum_{i<k}(2- U_{\alpha_{1}...
\alpha_{i-1}\alpha_{i+1}...\alpha_{n+1}}(\vec r) -
U_{\alpha_{1}...\alpha_{i-1}\alpha_{i+1}...
\alpha_{n+1}}(\vec r-\vec e_{\alpha_{i}}))$$
$$(2- U_{\alpha_{1}...
\alpha_{k-1}\alpha_{k+1}...\alpha_{n+1}}(\vec r) -
U_{\alpha_{1}...\alpha_{k-1}\alpha_{k+1}...
\alpha_{n+1}}(\vec r-\vec e_{\alpha_{k}})),\eqno(20)$$
or

$$H_{\alpha_{1}...\alpha_{n+1}}(\vec r) =
\frac{(n+1)n}{2}-
\frac{n}{2}
\sum_{i}(U_{\alpha_{1}...
\alpha_{i-1}\alpha_{i+1}...\alpha_{n+1}}(\vec r) +
U_{\alpha_{1}...\alpha_{i-1}\alpha_{i+1}...
\alpha_{n+1}}(\vec r-\vec e_{\alpha_{i}}))$$
$$+ \frac{1}{4}
\sum_{i<k}(U_{\alpha_{1}...
\alpha_{i-1}\alpha_{i+1}...\alpha_{n+1}}(\vec r) +
U_{\alpha_{1}...\alpha_{i-1}\alpha_{i+1}...
\alpha_{n+1}}(\vec r-\vec e_{\alpha_{i}}))$$
$$(U_{\alpha_{1}...
\alpha_{k-1}\alpha_{k+1}...\alpha_{n+1}}(\vec r) +
U_{\alpha_{1}...\alpha_{k-1}\alpha_{k+1}...
\alpha_{n+1}}(\vec r-\vec e_{\alpha_{k}})).\eqno(21)$$

Again as before (7a,b) the hyperplaquette variables $U$ are not
independent. Because
the hyperedge $\Omega_{\alpha_{1}...\alpha_{n+1}}(\vec r)$
belongs totally to an even number, $2(n+1)$, of hyperplaquettes
$\Omega_{\alpha_{1}...\alpha_{k-
1}\alpha_{k+1}...\alpha_{n+1}}(\vec r)$
and $\Omega_{\alpha_{1}...\alpha_{k-1}\alpha_{k+1}...
\alpha_{n}}(\vec r-\vec e_{\alpha_{k}})$, the product of the
corresponding
hyperplaquette variables is always one. Thus one has
constraints
$$\prod_{k=1}^{n+1} U_{\alpha_{1}...
\alpha_{k-1}\alpha_{k+1}...\alpha_{n+1}}(\vec r)
U_{\alpha_{1}...\alpha_{k-1}\alpha_{k+1}...
\alpha_{n+1}}(\vec r-\vec e_{\alpha_{k}})=1. \eqno(7c)$$
To resolve constraints (7c), let us attach an Ising
spin $\sigma$ in the center of each $d-n+1$
dimensional hypercube
$\Omega_{\alpha_{1}...\alpha_{n-1}}(\vec r)$ .
Then there exist spin configurations
$\sigma$, so that the hyperplaquette variables $U$ obey
\cite{wegner}

$$U_{\alpha_{1}...\alpha_{n}}(\vec r) =
\prod_{k=1}^{n}\sigma_{\alpha_{1}...
\alpha_{k-1}\alpha_{k+1}...\alpha_{n}}(\vec r)
\sigma_{\alpha_{1}...\alpha_{k-1}\alpha_{k+1}...
\alpha_{n}}(\vec r-\vec e_{\alpha_{k}})\eqno(9c)$$
The $\sigma$'s are defined by this equations up to global
and /or local gauge transformations \cite{wegner}. Equations
(5b) together with (21) and (9c) completely solve the problem.

Now let us consider special cases.
For $n=1$ this yields the competing interaction
for Ising spins

$$H_{\alpha\beta}(\vec r) = 1-\frac{1}{2}
(U_{\alpha}(\vec r)+ U_{\alpha}(\vec r-\vec e_{\beta})
+U_{\beta}(\vec r)+ U_{\beta}(\vec r-\vec e_{\alpha}))$$
$$+\frac{1}{2}(U_{\alpha}(\vec r)+ U_{\alpha}(\vec r-\vec
e_{\beta})
(U_{\beta}(\vec r)+ U_{\beta}(\vec r-\vec e_{\alpha})),\eqno(22)$$
which yields

$$H_{\alpha\beta}(\vec r)=$$
$$1-\frac{1}{2}(\sigma(\vec r)\sigma(\vec r-\vec e_{\alpha})+
\sigma(\vec r-\vec e_{\beta})\sigma(\vec r-\vec e_{\alpha}-\vec
e_{\beta})+
\sigma(\vec r)\sigma(\vec r-\vec e_{\beta})+
\sigma(\vec r-\vec e_{\alpha})\sigma(\vec r-\vec e_{\alpha}-\vec
e_{\beta}))$$
$$+\frac{1}{2}(\sigma(\vec r-\vec e_{\alpha})\sigma(\vec r-\vec
e_{\beta})+
\sigma(\vec r)\sigma(\vec r-\vec e_{\alpha}-\vec e_{\beta})).
\eqno(23)$$
Thus the Hamiltonian reads

$$H= \sum_{\vec r}\{ \frac{d(d-1)}{2}- (d-1)\sum_{\alpha}
\sigma(\vec r)\sigma(\vec r-\vec e_{\alpha})$$
$$+\frac{1}{2}\sum_{\alpha < \beta}(\sigma(\vec r-\vec
e_{\alpha})
\sigma(\vec r-\vec e_{\beta})+
\sigma(\vec r)\sigma(\vec r-\vec e_{\alpha}-\vec e_{\beta})) \}.
\eqno(24)$$
and again this Hamiltonian represents a magnetic system with
competing coupling constants adjusted to
$$J_{ferromagnet} = 2(d-1)~J_{antiferromagnet}. \eqno(12c)$$

For $n=2$ one obtains the gauge spin system

$$H_{\alpha\beta\gamma}(\vec r)=3-(U_{\alpha\beta}(\vec r)+
U_{\alpha\beta}(\vec r-\vec e_{\gamma})$$
$$+U_{\alpha\gamma}(\vec r)+
U_{\alpha\gamma}(\vec r-\vec e_{\beta})+U_{\beta\gamma}(\vec
r)+
U_{\beta\gamma}(\vec r-\vec e_{\alpha}))$$
$$+\frac{1}{4}(U_{\alpha\beta}(\vec r)+
U_{\alpha\beta}(\vec r-\vec
e_{\gamma}))(U_{\alpha\gamma}(\vec
r)+
U_{\alpha\gamma}(\vec r-\vec e_{\beta}))$$
$$+\frac{1}{4}(U_{\alpha\beta}(\vec r)+
U_{\alpha\beta}(\vec r-\vec e_{\gamma}))(U_{\beta\gamma}(\vec
r)+
U_{\beta\gamma}(\vec r-\vec e_{\alpha}))$$
$$+\frac{1}{4}(U_{\alpha\gamma}(\vec r)+
U_{\alpha\gamma}(\vec r-\vec e_{\beta}))(U_{\beta\gamma}(\vec
r)+
U_{\beta\gamma}(\vec r-\vec e_{\alpha})) \eqno(25)$$
and

$$H= \sum_{\vec r}\{\frac{d(d-1)(d-2)}{2} -
2(d-2)\sum_{\alpha < \beta}U_{\alpha\beta}(\vec r)$$
$$+\frac{1}{4}\sum_{\alpha , \beta < \gamma}
(U_{\alpha\beta}(\vec r)+
U_{\alpha\beta}(\vec r-\vec
e_{\gamma}))(U_{\alpha\gamma}(\vec
r)+
U_{\alpha\gamma}(\vec r-\vec e_{\beta}))\}, \eqno(26)$$
where

$$U_{\alpha\beta}(\vec r) = \sigma_{\alpha}(\vec r)
\sigma_{\alpha}(\vec r-\vec e_{\beta})
\sigma_{\beta}(\vec r)\sigma_{\beta}(\vec r-\vec e_{\alpha})
\eqno(27)$$
and
$$J_{plaquettes} = 8(d-2)~J_{rt~plaquettes}. \eqno(12d)$$
Thus one has a "ferromagnetic" interaction on each elementary
plaquette and an "antiferromagnetic" one on double-plaquettes
which form a right angle.
\vspace{1cm}

\section{Comments}
\vspace{.5cm}

 Let us consider the case of one-dimensional hypersurfaces
$M_1$,~~$d-n=1$, that is random walks on the lattice,
with the property that the straight "motion" does not
cost an energy, any right angle turns cost $\Theta(\pi /2)$
and for self intersections the number of pairs of links meeting
under a right angle times $\Theta(\pi /2)$ has to be paid.
As it is easy to see from our general result, when we
change the dimension of the target space $R^{d}$, then
we also should change the index $n$. In the two-dimensional
case we have $n=d-1=1$ and the spin Hamiltonian of the type
(11),(24). This special case is of particular interest,
because it can be mapped \cite{kadanoff,wu} to the eight-
vertex model \cite{baxter} with the weights

$$\epsilon_{1}=\epsilon_{3}=\epsilon_{4}=0,$$
$$\epsilon_{5}=\epsilon_{6}=\epsilon_{7}=\epsilon_{8}=
\Theta(\pi /2).\eqno(28)$$.

This is not the exactly solvable zero-field Baxter model, since
$\epsilon_1 \not= \epsilon_2$.
When $d=3$, for this type of scale invariant random walks
we should consider gauge spin system with $n=2$ and so on.

The models with $n=1$ are BNNNI~(biaxial next-nearest-
neighbor Ising) models which are relatives of the
ANNNI~(anisotropic next-nearest-neighbor Ising) models
\cite{selke,domb,elliott,bak,fisher,sinai}
with specially adjusted coupling constants, the restrictions of
which are dictated by the string geometry. It seems that the model
with
$d=2, n=1$ does not show a phase-transition \cite{selke,landau}. For
$d=3$ this is a random surface model $M_2$, for $d=4$ a
membrane
model $M_3$. In four and higher dimensions the random surface
models $M_2$ are again gauge models.

\vspace{1cm}

{\Large{\bf Acknowledgements}}
\vspace{.5cm}

One of the authors
(G.K.S.) is grateful to E. Floratos, R. Flume,
H. Fritsche, N. Papanicolaou,
E. Paschos and T. Tomaras for fruitful discussions and
to W. Greiner for useful comments and hospitality
in Frankfurt University.

This work was supported in part by the Alexander von
Humboldt Foundation.

\vfill

\vfill
\newpage
\vspace{.5cm}

\begin{thebibliography}{99}

\bibitem{savvidy1}R.V. Ambartzumian, G.K. Savvidy, K.G. Savvidy
and
G.S. Sukiasian. Phys. Lett. B275 (1992) 99

\bibitem{savvidy2}G.K. Savvidy and K.G. Savvidy. String fine
tuning.
Preprint UFTP 302/1992, hep-th 9208041,Int. J. Mod. Phys. A
(1993)

\bibitem{savvidy3}G.K. Savvidy, K.G. Savvidy. Gonihedric String and
Asymptotic Freedom. Preprint UFTP 320/1992, hep-th 9301001

\bibitem{wegner}F.J. Wegner, J. Math. Phys.12 (1971) 2259

\bibitem{kadanoff} L.P. Kadanoff and F.J. Wegner, Phys. Rev. B4
(1971) 3989

\bibitem{wu}R.Y. Wu, Phys.Rev. B4 (1971) 2312

\bibitem{baxter} R.J.Baxter, Exactly solved models in statistical
mechanics. Academic Press, London 1982

\bibitem{selke}W. Selke, Physics Reports 170 (1988) 213

\bibitem{domb} C. Domb, Adv. Phys. 9 (1960) 149

\bibitem{elliott} R.J. Elliott, Phys. Rev.124 (1961) 346

\bibitem{bak} P. Bak and J. von Boehm, Phys. Rev. Lett. 42 (1980)
122

\bibitem{fisher}M.E. Fisher and W.Selke, Phys. Rev. Lett. 44(1980)
1502

\bibitem{sinai}E.I. Dinaburg and Ya.G. Sinai, Comm.Math.Phys.
98 (1985) 119

\bibitem{landau}D.P. Landau, K. Binder, Phys. Rev. B 31 (1985)
5946

\end{thebibliography}
\end{document}